\documentclass[tikz,12pt]{article}
\usepackage{mheckII}

\usepackage[T1]{fontenc}

\usepackage{tikz-feynman}

\usepackage{manfnt}
\usepackage{longtable}

\usepackage{fancybox}

\usepackage{multirow}

\usepackage{blkarray}

\usepackage{tikz} 
\usetikzlibrary{matrix}

\theoremstyle{theorem}

\newtheorem*{cla}{Claim}

\theoremstyle{definition}

\newtheorem{rem}{Remark}

\def\Dsl{\,\raise.15ex\hbox{/}\mkern-13.5mu D}
\def\dsl{\,\raise.25ex\hbox{/}\mkern-10.5mu \partial}

\title{Moduli spaces of Calabi-Yau $d$-folds
 \\
as gravitational-chiral instantons 
}

\authors{Sergio Cecotti\footnote{e-mail: {\tt cecotti@sissa.it}}\vskip 9pt

\centerline{SISSA, via Bonomea 265, I-34100 Trieste, ITALY}
}

\abstract{Motivated by the swampland program, we show that the Weil-Petersson geometry of the moduli space of
a Calabi-Yau manifold of complex dimension $d\leq4$ is a gravitational instanton (i.e.\! a finite-action solution of the Euclidean equations of motion of gravity with matter). More precisely, the moduli geometry of Calabi-Yau $d$-folds ($d\leq4$) describes
instantons  of (E)AdS Einstein gravity coupled to a standard  chiral model.

From the point of view of the low-energy physics of string/M-theory compactified on the Calabi-Yau $X$,
the various fields propagating on its moduli space are the couplings appearing in the effective Lagrangian 
$\mathscr{L}_\text{eff}$.
}

\begin{document}
\maketitle


\newpage

\hskip6.5cm\begin{minipage}{220pt}
\begin{scriptsize}
{\it Tis true without lying, certain \& most true.
That wch is below is like that wch is above \& that wch is above is like yt wch is below to do ye miracles of one only thing}\\
\phantom{mm}\textsc{The Emerald Tablet,} translation by Isaac Newton

\end{scriptsize}
\end{minipage}

\vskip 20pt

The swampland program \cite{Vaf,OoV}  (for reviews see \cite{Rev1,Rev2}) looks for a characterization of the effective field theories which arise
as low-energy limits of consistent theories of quantum gravity, separating them from the vast swampland of effective theories which ``look'' consistent from 
a low-energy perspective, but cannot be completed to a fully consistent theory of quantum gravity. 
The program has produced a dozen or so conjectural necessary conditions (the ``swampland conjectures'' \cite{Vaf,OoV,Rev1,Rev2}) that all effective theories of quantum gravity should
satisfy.
\medskip

Dually, there is an inverse-swampland procedure. If we know that a certain effective theory \emph{does} arise from quantum gravity, 
we may apply the swampland ideas to \emph{predict} properties of the model. Often such properties are
too fancy for anybody to have enough fantasy to guess them, and they escaped us when
we looked at these theories with pre-swampland eyes. With post-swampland insight we know better.
This short note illustrates a simple application of the inverse-swampland strategy. The result echos the opening quotation, which was a major inspiration for Newton in formulating his own consistent theory of gravity. 
\medskip

We focus on the Weil-Petersson (WP) geometry of moduli spaces of compact Calabi-Yau (CY) manifolds $X_d$ of complex dimension $d\leq 4$. These CY manifolds 
describe stable supersymmetric vacua in string/M-theory, and the quantum-consistent low-energy effective theories around these vacua
are captured by the geometry of their moduli spaces. Hence inverse-swampland may yield new insights on the geometry of CY moduli. 
\medskip

We claim\footnote{\ The sharp version of the \textbf{Claim} will be presented momentarily, after fixing the necessary notation.} that for $d\leq4$ the CY moduli geometry yields a finite-action solution
to the classical equations of motion of a moduli-space field theory of the form
\be\label{action}
\int_M d^{\mspace{2mu}2m}x\;\sqrt{\det G}\left(-\frac{1}{2\kappa^2}\,R+ \frac{1}{2}\,G^{\alpha\beta}\,h(\phi)_{ab}\, \partial_\alpha \phi^a\,\partial_\beta \phi^b+\Lambda\right),\tag{1}
\ee
where $M$ is the CY moduli space, $G_{\alpha\beta}$ its metric, and $R$ its scalar curvature. In $d=1$ the only CY spaces are the elliptic curves; their
 moduli space has real dimension 2, so in this case the Einstein term in \eqref{action}
is topological, while the cosmological constant $\Lambda$ vanishes --  eqn.\eqref{action} reduces for $d=1$ to the classical Polyakov action of 
a string moving in the appropriate target space (which is also 2-dimensional). When $d=2$ the moduli-space Newton constant $\kappa^2$ is an adjustable parameter (this freedom reflects the fact that the moduli metric is always Einstein for CY 2-folds).
For $d=3,4$ the Newton constant
depends only on the dimension $d$ of the CY,
while the cosmological constant $\Lambda$ depends on $d$ and the complex dimension $m$ of the moduli space:
\be\label{Wconstants}
\frac{1}{\kappa^2}=\begin{cases} \text{free parameter} & d=2\\
\gcd(d,2) & d=3,4 
\end{cases},\qquad \Lambda=-\Big(m-1\Big)\!\left(\frac{m}{\kappa^2}+d\right).\tag{2}
\ee
When $m=1$ the real dimension of the moduli space $M$ is 2 and $\Lambda=0$, and again \eqref{action} reduces to the classical Polyakov action.
For $m>1$ the moduli-space gravity is ``dynamical'', and the cosmological constant negative. For $m=2$ we get instantons of a ``realistic''
four-dimensional gravity with matter. In the $d=2$ case 
the matter decouples in the limit $\kappa^2\to0$, and the moduli-space Einstein equations reduce to $R_{\alpha\beta}=-m\,G_{\alpha\beta}$. In this limit it is clear that the finiteness of the moduli volume
(one of the swampland conjectures \cite{OoV,Rev1,Rev2})
should be really understood as a \emph{finite action} condition for the moduli-space field theory \eqref{action}. This observation applies in general.

The matter part of the action \eqref{action} is a standard $\sigma$-model with target a locally symmetric space
$\Gamma\backslash G(\R)/K$, where the non-compact real Lie group $G\equiv G(\R)$ is the automorphism group of the corresponding Griffiths period domain, that is, explicitly
\be\label{groups}
G(\R)= \left\{\begin{aligned}&Sp(2n,\R) &&\text{with }2n=\sum_{p=0}^d h^{p,d-p}_\mathrm{prim} &&\text{for $d$ odd}\\
&SO(s,t) &&\text{with }s=\sum_{k=0}^d h^{2k,d-2k}_\mathrm{prim},\  \  t=\sum_{k=0}^1 h^{2k+1,d-1-2k}_\mathrm{prim}&&\text{for $d$ even,}\end{aligned}\right.\tag{3}
\ee
where $\{h^{p,d-p}_\text{prim}\}$ are the primitive Hodge numbers in middle dimension\footnote{\ $\mathbb{H}^{p,q}(X_d)$ stands for the space of harmonic forms of type $(p,q)$ on $X_d$.}
\be
h^{p,d-p}_\text{prim}\overset{\rm def}{=}\dim_\C\Big\{ \xi\in \mathbb{H}^{p,d-p}(X_d)\colon \omega\wedge\xi=0\Big\}\qquad \omega\equiv\text{K\"ahler form.}\tag{4}
\ee
$K\subset G(\R)$ is a maximal compact subgroup. For comparison, we recall that the Griffiths period domain $D$  is the reductive  coset $G(\R)/H$ where \cite{Gbook,deligne,periods}
\be\label{WH}
H = SO(h^{d/2,d/2}_\text{prim})\times \prod_{0\leq p<d/2} U(h^{p,d-p}_\text{prim})\subset K.\tag{5}
\ee
One may replace $G(\R)$ with the Lie subgroup $MT(\R)\subseteq G(\R)$ given by the real locus of the Mumford-Tate group $MT$ \cite{reva,revb,MT4} of the moduli of $X_d$.
Indeed the relevant matter field configuration $\phi\colon M\to \Gamma\backslash\G(\R)/K$ has image in the totally geodesic submanifold\,\footnote{\ This statement follows from the \textbf{structure theorem} for the period map \cite{reva,revb,MT4}.} 
\be
\Gamma\backslash MT(\R)/[MT(\R)\cap K]\hookrightarrow \Gamma\backslash G(\R)/K.\tag{6}
\ee
The discrete group $\Gamma\subset MT(\Z)\subseteq G(\Z)$ is the monodromy group of the CY period map.
\medskip

Very roughly speaking, the moduli space has the form $M=\cg\backslash \widetilde{M}$, with $\widetilde{M}$ diffeomorphic to $\R^{2m}$
and $\cg$ a discrete group with a  neat subgroup of finite index. We call $\cg$ the $U$-duality group. (In the present context it is isomorphic to the monodromy group $\Gamma$, but we denote them with distinct symbols for clarity).
\medskip

A classical solution to \eqref{action} consists of two pieces of data: a metric $G_{\alpha\beta}$ on $\widetilde{M}$ admitting $\cg$ as a group of isometries,
 and a harmonic map $\tilde\phi\colon \widetilde{M}\to G(\R)/K$ satisfying the $\cg$-equivariant condition
 \be\label{twisted1}
 \tilde\phi(g\cdot x)= \rho(g)\cdot \tilde\phi(x),\qquad x\in \widetilde{M},\quad g\in\cg\tag{7}
 \ee
 for some group homomorphism
 \be\label{twisted2}
 \rho\colon\cg\twoheadrightarrow \Gamma\equiv \rho(\cg)\subset G(\Z)\subset G(\R)\tag{8}
 \ee
 called the monodromy representation. When \eqref{twisted1} holds, one also says that the map $\tilde\phi$ is \emph{twisted} by $\rho$. A $\rho$-twisted map descends to a map $\phi\colon M\to \Gamma\backslash G(\R)/K$,
 and we shall use $\tilde\phi$ and $\phi$ interchangeably.
 Our claim states that the CY moduli geometry is given by a pair
 $(G_{\alpha\beta},\tilde\phi)$ which satisfies the equations of motion following from the action \eqref{action}:
 \begin{align}\label{torsionless}
&D^\alpha\mspace{1mu}\partial_\alpha\tilde\phi=0\tag{9}\\
&R_{\alpha\beta}-\frac{1}{2}\,G_{\alpha\beta}\,R+\kappa^2\mspace{2mu}\Lambda\, G_{\alpha\beta}= \kappa^2\, T_{\alpha\beta},\tag{10}\label{eisteinequation}
\end{align}
where the derivative $D_\alpha$ is covariant for the combined Levi-Civita connections of $T^*\mspace{-3mu}M$ and $\tilde\phi^*T(G/K)$.
Eqn.\eqref{torsionless} just expresses the fact that $\tilde\phi$ is a \emph{harmonic} map $\widetilde{M}\to G(\R)/K$
 for the source-space metric $G_{\alpha\beta}$.
\medskip

Having fixed the notation, let us make our claim sharp:

\begin{cla}\label{ccccla} Let $G_{\alpha\beta}$ be the Weil-Petersson metric on the complex moduli space $M$ of a CY $d$-fold,
with $d\leq4$, and let $\phi$ be the composed map $\phi\equiv \pi\circ p$ where
\begin{align}
&p\colon M\to \Gamma\backslash G(\R)/H &&\text{\rm the (global) Griffiths period map \cite{Gbook,deligne,periods,reva,revb,MT4}}\label{pre1}\tag{11}\\
&\pi\colon \Gamma\backslash G(\R)/H\twoheadrightarrow \Gamma\backslash G(\R)/K &&\text{\rm the canonical projection.}\label{pre2}\tag{12}
\end{align}
Then the pair $(G_{\alpha\beta},\phi)$ is a finite-action solution to eqns.\eqref{torsionless},\eqref{eisteinequation}
with constants as in eqn.\eqref{Wconstants}.
\end{cla}

\begin{rem} Eqn.\eqref{torsionless} remains true when
 $d\geq5$. The WP metric $G_{\alpha\beta}$ still satisfies an ``Einstein-like'' equation.
 However it seems that one cannot construct an \emph{off-shell} action with \emph{positive kinetic terms}
 whose canonical energy-momentum tensor yields the source term in the equation. This is to be expected since the moduli geometry of $d\geq5$ Calabi-Yau's is not required to have ``magical'' properties by swampland consistency conditions. 
\end{rem}

\subsubsection*{The moduli space fields $(G_{\alpha\beta},\phi)$ as effective couplings}

To make explicit contact with the swampland program, let us recall the low-energy 4d effective Lagrangian of Type IIB compactified
on the (simply-connected) CY 3-fold $X_3$ 
 \begin{equation}\label{lag}
\mathscr{L}_\text{IIB}= \sqrt{-g}\Big(-\frac{1}{2}R+\frac{1}{2}G(\varphi)_{\alpha\beta}\,\partial^\mu \varphi^\alpha\partial_\mu \varphi^\beta +
\frac{i}{16\pi}\tau(\varphi)_{ab}F^a_+F^b_+-\frac{i}{16\pi}\bar\tau(\varphi)_{ab}F^a_-F^b_-+\cdots\Big),\tag{13}
\end{equation}
where for brevity we wrote only the matter terms involving the bosonic fields of the vector-multiplets. 
The metric $G_{\alpha\beta}$ appearing in the vector-multiplet scalars' kinetic terms coincides with the Weil-Petersson metric on
the moduli space $M$ of $X_3$ \cite{Cecotti:1989kn,stro}. For a fixed point $\varphi\in \widetilde{M}$, the gauge
coupling $\tau(\varphi)_{ab}$ is a symmetric complex matrix with positive imaginary part, that is, a point in the Siegel upper half-space
\be
\tau(\varphi)_{ab}\in Sp(2h^{2,1}+2,\R)\big/U(h^{2,1}+1)\equiv G(\R)/K\qquad \big(\text{cfr.\! eqn.\eqref{groups} with $d=3$}\big).\tag{14}
\ee 
Hence the 4d gauge coupling may be identified with the map
\be
\tilde\phi\colon \widetilde{M}\to G(\R)/K,\qquad \tilde\phi\colon \varphi\mapsto \tau(\varphi)_{ab}.\tag{15}
\ee
However this way of describing the gauge couplings is \emph{not} intrinsic, since $\tau(\varphi)_{ab}$ depends on a choice of duality frame.
Even worse: the 
$\tau(\varphi)_{ab}$ are \emph{multi-valued}\,\footnote{\ The \textbf{structure theorem} yields the dichotomy: either the gauge couplings $\tau(\varphi)_{ab}$
are field-independent numerical constants (as in the case of rigid CY 3-folds \cite{Cecotti:2018ufg}) or $\tau(\varphi)_{ab}$ \emph{must be}
multivalued.} functions on $M$ because when
we go around a non-trivial loop in $M$ we come back with a rotated electro-magnetic duality frame. The intrinsic description of the gauge couplings is instead given by the quotient map
\be\label{intrinsic}
\phi\colon M\equiv \cg\backslash \widetilde{M}\to \Gamma\backslash G(\R)/K,\qquad \phi\colon [\varphi]\mapsto [\tau(\varphi)_{ab}].\tag{16}
\ee
In other words, the lifted gauge coupling map $\tilde\phi$ is twisted by the monodromy representation $\rho$ as in eqns.\eqref{twisted1},\eqref{twisted2}. Indeed, the $U$-duality group $\cg$ acts both on $\widetilde{M}$ (by isometries) and on the vector field-strengths (by electro-magnetic dualities)
while leaving the physical energy-momentum tensor $T_{\mu\nu}$ invariant; this entails that the `naive' gauge coupling map
$\tilde\phi$ is twisted by the monodromy representation $\rho$ of $\cg$.
This being understood, the on-shell configurations \eqref{pre1},\eqref{pre2} of the two fields $(G_{\alpha\beta},\phi)$  which propagate in the moduli space $M$ are \emph{exactly} the same as the 
 couplings appearing in the Type IIB 4d effective Lagrangian $\mathscr{L}_\text{IIB}$:
\be
\begin{split}
G_{\alpha\beta}\ &\equiv\  \text{vector-multiplet scalars' metric}\\
\phi\  &\equiv\  
\text{(intrinsically-defined) gauge couplings, eqn.\eqref{intrinsic}.}
\end{split}\tag{17}
\ee

\begin{rem} More generally, it is pretty obvious that \emph{all} couplings appearing in the Lagrangian $\mathscr{L}_\text{eff}$ of \emph{any} 4d supergravity
which is consistent with the swampland conjectures \cite{Vaf,OoV,Rev1,Rev2} and has $\geq8$ supercharges
describe (as functions of the scalar fields) gravitational instantons. Again, the finite volume conjecture gets re-interpreted
as the statement that the field configuration in moduli space which decribes the effective couplings in $\mathscr{L}_\text{eff}$
 has \emph{finite action.}
\end{rem}

\begin{rem} Conversely, the $\cn\geq2$ \textsc{sugra}'s which do \emph{not} satisfy the swampland conjectures are not
gravitational instantons.
E.g.\! the homogeneous $\cn=2$ models constructed in \cite{Cecotti:1988ad}, all of which fall in the swampland \cite{swampIII},
do satisfy eqns.\eqref{torsionless},\eqref{eisteinequation} but have \emph{infinite} action.
\end{rem}
%


\subsection*{Details and proofs}

Of course, once we have strong reasons -- such as the swampland story -- to believe that something
\emph{ought to} be true, we look for actual proofs rather than relying on widely believed conjectures. 
Our treatment in this note will be totally rigorous (except that we do not discuss the singularities of the
relevant solution -- a crucial issue, but not one consistent with the purpose of writing a \emph{short} note).
\medskip

We present an informal discussion of the general picture in \S.\,\textbf{0}. Then in \S\S.\textbf{1}-\textbf{10} we enter in the technical details, and
write explicit expressions for all relevant quantities. 

\subsubsection*{0. An informal sketch}

There are two approaches (or languages) for the geometry of Calabi-Yau moduli spaces:
\textit{(i)} Griffiths theory of variations of Hodge structures (VHS) \cite{Gbook,deligne,periods}, and \textit{(ii)} $tt^*$ geometry \cite{Cecotti:1991me,Cecotti:1992rm,Cecotti:2013mba}. Equivalence of the two viewpoints (in the appropriate contexts) was 
proven in \cite{Cecotti:1990wz}\!\!\cite{Cecotti:1991me} (and enshrined in the math literature as a theorem in \cite{simpson}).  We shall use both languages, with a preference for the second one.
\medskip

On the moduli space of a Calabi-Yau $d$-fold there is an infinite family of \emph{a priori} distinct canonical K\"ahler metrics.
From the point of view of VHS this plethora arises because the Griffiths period domain $D\equiv G(\R)/H$ \cite{Gbook,deligne,periods}
carries several holomorphic homogeneous line bundles whose canonical connection has a curvature which is positive when
restricted to the Griffiths horizontal tangent bundle. The pull-back to $M$, \emph{via} the period map $p\colon M\to \Gamma\backslash D$,
of any one of these curvatures yields a K\"ahler form on $M$. There is one horizontally-positive line-bundle which exists on the period domain $D$
for all Hodge numbers $\{h^{p,q}\}$, namely the Griffiths canonical line bundle \cite{griIII}.
The corresponding K\"ahler metric is called the \emph{Hodge metric} $K_{j\bar k}$, and is the best behaved one in the family.
In the case of Calabi-Yau $d$-folds, one has $h^{d,0}=1$ and there is another important horizontally-positive line bundle whose sections are the holomorphic $(d,0)$-forms.
Its curvature defines the \emph{Weil-Petersson} (WP) K\"ahler metric $G_{j\bar k}$.  $K_{j\bar k}$, $G_{j\bar k}$ do not exhaust the list of canonical VHS metrics. Taking linear combinations with positive coefficients of the several canonical metrics,
 we construct a \emph{convex cone} $\mathscr{C}_d$ of God-given K\"ahler metrics on the moduli $M$.
The term ``God-given'' here has a precise technical meaning:
$$
\text{\begin{minipage}{270pt}\it the $U$-duality group $\cg$ acts by isometries with respect to {\textbf{all}} K\"ahler metrics in the convex cone  $\mathscr{C}_d$.\end{minipage}}
$$
This is quite remarkable, since $\cg$ is a ``huge'' group: for CY 3-folds, say,
unless the IIB 4d gauge couplings $\tau(\varphi)_{ab}$ are numerical constants (i.e.\! the CY is rigid \cite{Cecotti:2018ufg}), the Zariski closure of $\cg$ is a semi-simple real Lie group of positive dimension \cite{swampIII}. We stress that $\mathscr{C}_d$ is a cone of actual K\"ahler \emph{metrics,} not just K\"ahler classes. We write $K^{(c)}_{j\bar k}$ for the K\"ahler metric associated to a point $c\in\mathscr{C}_q$.
 
Not all the canonical K\"ahler metrics are independent: decreasing the dimension $d$ of the Calabi-Yau space,
the dimension of the cone gets smaller and smaller
\be
\begin{split}
\dim\mathscr{C}_d &= \mathrm{rank}\mspace{2mu}\Big(\text{group of \emph{homogeneous} line-bundles over $G/H$}\Big)\equiv\\
&\equiv \mathrm{rank}\,\mathsf{Hom}(H,U(1)) \leq \left[\frac{d+1}{2}\right]\qquad\qquad \begin{smallmatrix}\text{equality for \textit{connected}}\\
\text{Hodge structures\phantom{mmn}}
\end{smallmatrix}
\end{split}\tag{18}
\ee 
For $d\leq 4$ the inequality is saturated provided the CY is not rigid.
\medskip

The Ricci tensor $R_{j\bar k}$ of the WP metric
$G_{j\bar k}$ is also the pull-back of the curvature of a homogenous line-bundle on $D$, hence $R_{j\bar k}$ belongs to the linear span of the God-given metrics, i.e.\! $R_{j\bar k}$ can be written as a linear combination of
canonical VHS metrics. This is consistent since
all K\"ahler metrics $K^{(c)}_{j\bar k}$ 
satisfy the same Bianchi identity as the WP Ricci tensor
\be\label{Bianchi}
D^j\mspace{-5mu}\left(K^{(c)}_{j\bar k}-\frac{1}{2}\, G_{j\bar k}\, K^{(c)}\right)=0,\qquad \text{where }K^{(c)}\overset{\rm def}{=} 2\,G^{j\bar k}K^{(c)}_{j\bar k},\tag{19} 
\ee
and $D_j$ is the Levi-Civita connection of the WP metric.
On the moduli space of CY $d$-folds we have an identity of the form
\be\label{EEE}
R_{j\bar k} -\frac{1}{2}\,G_{j\bar k}\,R =\sum_{c=1}^{\dim\mathscr{C}_d} \lambda^{(c)}\!\left(K^{(c)}_{j\bar k}-\frac{1}{2} G_{j\bar k}\, K^{(c)}\right)\tag{20}
\ee
for certain numerical constants $\lambda^{(c)}$ (to be computed in \S.\,\textbf{10} below). We conclude that the WP metric $G_{j\bar k}$ is a solution to the Einstein equations \emph{provided} the \textsc{rhs} of \eqref{EEE}
may be written as a physically sound energy-momentum tensor plus a cosmological constant term.
\medskip

 For CY's of dimension $d=1,2$ one has $\dim \mathscr{C}_d=1$, so all
canonical VHS metrics are multiples of the WP one which then must be Einstein, i.e.\! $R_{j\bar k}=-\Lambda\, G_{j\bar k}$ for some $\Lambda$. 
\medskip

For $d=3,4$ there is a 2-parameter family of VHS K\"ahler metrics on $M$,
and then the Ricci tensor of the WP metric must be a linear combination of the WP metric $G_{j\bar k}$ and the Hodge one
$K_{j\bar k}$. Now our \textbf{Claim} follows from the fact\footnote{\ See eqn.\eqref{WT} below.} that the combination entering in \eqref{EEE}
\be
K_{j\bar k}-\frac{1}{2}\,G_{j\bar k}\,K\equiv T_{j\bar k},\tag{21}
\ee
is nothing else than the canonical energy-momentum tensor of the $\Gamma\backslash G(\R)/K$ $\sigma$-model evaluated
on the \emph{on-shell} field configuration \eqref{pre1},\eqref{pre2}.
\medskip

Having sketched the general picture, let us now write the explicit formulae.

\subsubsection*{1. Review of the $\Gamma\backslash G/K$ $\sigma$-model} 

$G\equiv G(\R)$ is a non-compact real Lie group and $K\subset G$
a maximal compact subgroup which is fixed by the Cartan involution $\theta$.
A map $M\to \Gamma\backslash G/K$ can be lifted (non-uniquely) to a map 
$E\colon \widetilde{M}\to G$. We see  the lifted map as a \emph{field} $E(x)$ on the Euclidean space-time $\widetilde{M}$
which takes value in the concrete matrix group given by the Hodge representation of $G(\R)$ which, for the groups
in \eqref{groups}, is the fundamental one ($\boldsymbol{2n}$ and, respectively, $\boldsymbol{s+t}$).
The field $E(x)$ is twisted by the monodromy representation (cfr.\! eqn.\eqref{twisted1})
\be
E(g\cdot x)=\rho(g)\cdot E(x),\qquad \forall\; g\in \cg,\tag{22}
\ee
so it descends to a field (or map) $\underline{E}\colon M\to \Gamma\backslash G$. Two field configurations,
$E(x)$ and $E(x)^\prime$, which differ by the multiplication on the right by
a position-dependent element of $K$, are declared to be gauge-equivalent (i.e.\! the same physical configuration)
\be
E(x)^\prime \sim E(x)\quad\Longleftrightarrow\quad E(x)^\prime=E(x)\, U(x),\quad \text{with }U(x)\in K.\tag{23}
\ee 
By a global field configuration $E(x)$ we actually mean a local lift $E(x)_\alpha\colon U_\alpha\to G$
for each open set of a cover $\cup_\alpha U_\alpha=\widetilde{M}$, with transition function $E(x)_{\beta}^{-1}E(x)_\alpha\in K$
in $U_\alpha\cap U_\beta$.

We have the $K$-principal bundle $\varpi\colon G\to G/K$, and the physical gauge-invariant map is
$\varpi\circ E\colon \widetilde{M}\to G/K$ or, more precisely, its $\cg$-equivariant quotient 
\be
\underline{\varpi}\circ \underline{E}\colon M\to \Gamma\backslash G/K.\tag{24}
\ee

 We adopt the following notation:
for $a\in\mathfrak{g}$ (the Lie algebra of $G$), $a^{\rm e}$ and $a^{\rm o}$ denote, respectively, the projection on the even and odd parts under the Cartan involution $\theta$. The action of the $\sigma$-model with target space $\Gamma\backslash G/K$
is 
\be\label{sigmamode}
\frac{1}{2}\int_M d^{\mspace{1.7mu}2m} x\,\sqrt{\det G}\;G^{\alpha\beta}\,\mathrm{tr}\mspace{-1mu}\Big[(E^{-1}\partial_\alpha E)^{\rm o}(E^{-1}\partial_\beta E)^{\rm o}\Big]\tag{25}
\ee
(one checks that it is $K$ gauge invariant). The energy-momentum tensor is
\be\label{emtens}
T_{\alpha\beta}= \mathrm{tr}\mspace{-1mu}\Big[(E^{-1}\partial_\alpha E)^{\rm o}(E^{-1}\partial_\beta E)^{\rm o}\Big]-\frac{1}{2}\,
G_{\alpha\beta}\, G^{\gamma\delta}\,\mathrm{tr}\mspace{-1mu}\Big[(E^{-1}\partial_\gamma E)^{\rm o}(E^{-1}\partial_\delta E)^{\rm o}\Big].\tag{26}
\ee
The equations of motion say (by definition) that the field $E$ is on-shell if and only if the corresponding physical map,
$\underline{\varpi}\circ \underline{E}\colon M\to\Gamma\backslash G/K$, is \emph{harmonic.}

\subsubsection*{2. Pluri-harmonic maps}

 Let $M$ be a K\"ahler manifold and $Y$ any Riemannian manifold.
A map $f\colon M\to Y$ is \emph{pluri-harmonic} iff 
\be\label{plurihha}
D_j\partial_{\bar k}f=0.\tag{27}
\ee
Note that the covariant derivative $D_j$ contains only the
Levi-Civita connection of $f^*TY$. If $f$ is pluri-harmonic, $G^{j\bar k}D_j\partial_{\bar k}f\equiv 0$, so $f$ is \emph{a fortiori}
harmonic, hence a classical solution of the $\sigma$-model with target space $Y$ and source space $M$.
We stress that \eqref{plurihha} does not contain the K\"ahler metric of $M$, so a pluri-harmonic map is harmonic for \emph{all}
choices of K\"ahler metric.
\medskip

In our application $Y$ is the locally symmetric space $\Gamma\backslash G/K$
which is non-compact of finite volume.\footnote{\ More precisely: we may reduce to the case of finite volume without loss of generality.
If $\Gamma$ is thin, replace it by an arithmetic group which contains it.} We assume $M$ to be non-compact and the existence of \emph{some} complete K\"ahler metric $\mathring{g}_{j\bar k}$ on $M$ of finite volume; \emph{``some''} means that the reference metric $\mathring{g}_{j\bar k}$ may have nothing to do with the 
physical metric $G_{j\bar k}$.\footnote{\ Actually, taking as reference metric $\mathring{g}_{j\bar k}$ the physical one  $G_{j\bar k}$
only improves the situation.} We claim that in these circumstances\,\footnote{\ For the statement to be true, one needs to require some extra ``regularity''
conditions which are tautologically satisfied for the spaces of interest.} any classical solution $E_0$ of the $\sigma$-model (defined with the space-time metric
$\mathring{g}_{j\bar k}$) which has \emph{finite action}
\be
S(E_0)\equiv \int_M d^{\mspace{1.5mu}2m}x\,(\det\mathring{g})\;\mathring{g}^{j\bar k}\,\mathrm{tr}\mspace{-1mu}\Big[(E_0^{-1}\partial_j E_0)^{\rm o}(E_0^{-1}\partial_{\bar k} E_0)^{\rm o}\Big]\tag{28}
 <\infty
\ee
is automatically pluri-harmonic, hence a solution of the equations of motion for \emph{any} other choice of K\"ahler metric $G_{j\bar k}\neq \mathring{g}_{j\bar k}$
on the source space $M$. The claim is a special case of a more deep fact, crucial for the swampland story, which shall be discussed elsewhere in its proper context. A sketch of the proof (for the special situation at hand) will be given in the next paragraph
after the introduction of the necessary notation.

\subsubsection*{3. Review of $tt^*$ geometry}

A $tt^*$ geometry on the complex manifold $M$ is just a pluri-harmonic map $M\to \Gamma\backslash G/K$
of finite action,\footnote{\ In this note we add to the ``standard'' definition of $tt^*$ geometry the condition that the underlying pluri-harmonic map has \textit{finite action;} all $tt^*$ geometries arising from physics satisfy this condition. For the geometries relevant for the present paper this will be shown in \S.\,5 below.} 
where $G$, $K$ and $\Gamma$ are as in \S.\textbf{1}. 
As in that paragraph, the $tt^*$ map may be lifted to a map $S\colon \widetilde{M}\to G$. Again, we see $S$
as a field on $\widetilde{M}$ taking value in the concrete matrix group $G$. In facts, $S$ is just a special instance of
the $\sigma$-model field $E$ of \S.\textbf{1}: $S$ is not just an on-shell field configuration, it satisfies the stronger condition of being pluri-harmonic
(this is essentially automatic in the present circumstances, see below). 
$S$ has a direct physical meaning: in the $tt^*$ literature \cite{branes}\!\!\cite{Cecotti:2013mba,Cecotti:2014wea}
$S$ is called the \emph{BPS brane amplitude} (for some value $\zeta=e^{i\theta}$ of the spectral parameter which depends on the chosen lift).
\medskip

Since $M$ is complex, we may decompose the differential forms into definite type 
\be\label{eopro}
(S^{-1}dS)^{\rm e}=A+\bar A, \qquad (S^{-1}dS)^{\rm o}=C+\bar C,\tag{29}
\ee
where unbarred (barred) stands for type (1,0) (resp.\! (0,1)). We introduce the $K$-covariant Dolbeault differentials
$D=\partial+A$ and $\bar D=\bar\partial+\bar A$. We have the identity
\be\label{TT1}
DC= \big((d+S^{-1}dS)^2\big)^{\mspace{-2mu}\rm o}\Big|_{(2,0)\text{-part}}=0,\tag{30}
\ee
while the condition that the $\cg$-twisted $tt^*$ map $\varpi\circ S\colon\widetilde{M}\to G/K$ is pluri-harmonic reads 
\be\label{TT2}
\bar D C=0.\tag{31}
\ee

Before proceeding, let us pause a while to sketch the idea behind the claim made at the end of the previous paragraph:
if $M$ is a non-compact K\"ahler manifold of finite volume,
satisfying some mild regularity condition, then a harmonic map 
$M\to \Gamma\backslash G/K$ of finite-energy is automatically pluri-harmonic.
One starts
from Simpson's Bochner-formula \cite{simpson,simp2} for harmonic maps with source space a K\"ahler manifold $M$.
In the present set-up and notations, this Bochner-formula takes the form (see eqns.(4.1)-(4.8) of \cite{swampIII})
\be\label{boch}
\varpi\circ S\ \text{harmonic}\quad\Rightarrow\quad
\overline{D}^i \overline{D}^j\mathrm{tr}(C_iC_j)=\|\overline{D}C\|^2+\text{non-negative}.\tag{32}
\ee
When $M$ is compact, the integral of the total derivative in the \textsc{lhs} vanishes, so the two non-negative terms in the \textsc{rhs}
should vanish separately, and we have
$\overline{D}C=0$, which is the statement that the map $\varpi\circ S$ is \emph{pluri}-harmonic. If $M$ is \emph{non}-compact,
the integral of the \textsc{lhs} yields a boundary term at infinity, and the same conclusion
applies \emph{provided} we can show that this boundary term vanishes. This vanishing condition at infinity holds when $M$ has finite volume and satisfies the mild technical assumptions (roughly: ``the ends of $M$ at $\infty$ have vanishing size'') while
the harmonic map has finite-energy (i.e.\! it is ``trivial at $\infty$'').
\medskip

We return to $tt^*$ geometry. A short computation \cite{dubrovin}\!\!\!\cite{swampIII} shows that the compatibility condition of  
 \eqref{TT1} with \eqref{TT2}, 
$[D,\bar D]C=0$, implies 
\be\label{TT3}
C\wedge C=0,\tag{33}
\ee 
which expresses the fact that the $tt^*$ chiral ring $\mathscr{R}$ is commutative.  
 Using \eqref{TT1},\eqref{TT2}, and \eqref{TT3}
one checks that the Maurier-Cartan identity $(d+S^{-1}dS)^2=0$ is equivalent to the statement that
the\footnote{\ Here and below $\mathfrak{g}^\C\equiv \mathfrak{g}\otimes\C$, where $\mathfrak{g}$ is the Lie algebra of the real Lie group $G$.} $\mathfrak{g}^\C$-valued connection 
\be\label{lax}
\boldsymbol{\nabla}^{(\zeta)}\overset{\rm def}{=}d+A+\bar A+ \zeta\, C+\zeta^{-1}\, \bar C\tag{34}
\ee  
is \emph{flat} for all values of the spectral parameter $\zeta\in\mathbb{P}^1$
\be\label{lax2}
\big(\boldsymbol{\nabla}^{(\zeta)}\big)^2\equiv 0.\tag{35}
\ee 
Eqn.\eqref{lax2} is the Lax form of the $tt^*$ PDEs
\cite{Cecotti:1991me,Cecotti:1992rm,Cecotti:2013mba,dubrovin}.
\medskip

The application of $tt^*$ geometry to 2d (2,2) QFT \cite{Cecotti:1991me} works as follows. Let $P$ be the complex space of
 $F$-term\footnote{\ The distinction between $F$-term and \emph{twisted} $F$-term is a matter of convention.
We loosely say ``$F$-term'' to mean either one, depending on the particular application one has in mind.} parameters. Over $P$
we have the vector bundle $\mathscr{V}$ whose fiber at $p\in P$ is the space of \textsc{susy}
vacua of the QFT with couplings $p$. The $tt^*$ connection $D+\bar D$ acts on $\mathscr{V}$; by eqn.\eqref{lax2} it endows $\mathscr{V}$ with a holomorphic structure.
By construction the $tt^*$ connection is metric for the QFT Hilbert space inner product, and hence it is the unique Chern connection on $\mathscr{V}$
(and also the Berry one). In a holomorphic gauge we have 
\be\label{hologauge}
A=g\mspace{1.3mu}\partial\mspace{0.6mu} g^{-1},\qquad \bar A=0,\tag{36}
\ee
where $g\equiv (g_{a\bar b})$ is the Hilbert space Hermitian metric along the fibers (the $tt^*$ metric \cite{Cecotti:1991me}).
The spectral flow isomorphism \cite{chiralr} states that 
\be\label{isoiso}
\mathscr{V}\cong \mathscr{R}\hookrightarrow\mathrm{End}(\mathscr{V})\cong
\mathscr{V}\otimes\mathscr{V}^\vee \cong \mathscr{V}^{\otimes 2},\tag{37}
\ee
where $\mathscr{R}$ is the holomorphic bundle whose fiber $\mathscr{R}_p$ is the chiral ring at $p\in P$. The last isomorphism
in \eqref{isoiso} is the reality structure\footnote{\ Equivalently, the topological metric $\eta$ \cite{dubro1}.} \cite{Cecotti:1991me}.
The holomorphic vacuum bundle $\mathscr{V}$ is then isomorphic to a sub-bundle of its tensor-square $\mathscr{V}^{\otimes 2}$.
This yields an induction on the bundle metrics: start with the fiber metric $g$ for $\mathscr{V}$; it
 induces a fiber metric for $\mathscr{V}^{\otimes 2}$, and its restriction to the sub-bundle
 $\mathscr{R}$ is then a second fiber metric $h$ for $\mathscr{V}$ (one may iterate the process \emph{ad infinitum}).

\subsubsection*{4. Superconformal $tt^*$ geometry}

The discussion in \S.\,\textbf{3} applies to all 2d (2,2) QFTs \cite{Cecotti:1991me}.
 When the (2,2) QFT is superconformal one is mainly interested in the $tt^*$ geometry restricted to the conformal submanifold $M\subset P$
  of (exactly) marginal deformations.  When so restricted, the holomorphic bundles 
 $\mathscr{R}\to M$ and $\mathscr{V}\to M$ get graded by the superconformal $U(1)$ charge $q$
 \be\label{deedccc}
 \mathscr{R}=\bigoplus_{q=0}^{\hat c}\mathscr{R}_q,\qquad \mathscr{V}=\bigoplus_{q=-\hat c/2}^{\hat c/2}\mathscr{V}_q,\qquad \mathscr{V}^\vee_q\cong \mathscr{V}_{-q},\qquad
\mathscr{V}_{q-\hat c/2} \cong\mathscr{R}_q.\tag{38}
 \ee
The decomposition of $\mathscr{V}$ is orthogonal for the $tt^*$ fiber metric $g$ \cite{Cecotti:1991me}. 
Conformal perturbation theory gives us the isomorphism\;\footnote{\ In the VHS language this isomorphism is called the ``local Torelli theorem''.}
 \be\label{loctorelli}
\begin{smallmatrix}\text{{\it holomorphic} tangent bundle}\\
\text{of conformal manifold $M$\phantom{mn}}\end{smallmatrix}\qquad T\mspace{-1mu}M\cong \mathscr{V}_{1-\hat c/2}\cong \mathscr{R}_1\hookrightarrow \mathscr{V}^{\otimes2}.\qquad\phantom{mmmmm}\tag{39}
 \ee
The Hodge metric is the metric on $T\mspace{-1mu}M$ given by the induced metric on $\mathscr{R}_1$ as a sub-bundle of $\mathscr{V}^{\otimes 2}$,
while the WP metric is the \emph{normalized} $tt^*$ metric restricted to $\mathscr{V}_{1-\hat c/2}$ 
 \cite{Cecotti:1991me,Cecotti:1992vy}:
\be\label{xrqe}
\text{WP metric on $M$}=\frac{g|_{\mathscr{V}_{1-\hat c/2}}}{g_{-\hat c/2}},\qquad
\text{Hodge metric on $M$}=h|_{\mathscr{V}_{1-\hat c/2}}.\tag{40}
\ee
When the $tt^*$ geometry describes the complex moduli of a CY $d$-fold $X_d$ -- that is, when the 2d (2,2) SCFT is the $X_d$ $\sigma$-model -- one has $\hat c=d$ and 
\be\label{spectrum}
\mathrm{rank}\,\mathscr{V}_q= h^{d/2-q,\mspace{2mu}d/2+q}_\text{prim},\quad \text{in particular, }\mathrm{rank}\,\mathscr{V}_{\mp \hat c/2}=1.\tag{41}
\ee
In this case the $tt^*$ Lie group $G\equiv G(\R)$, introduced in \S.\,\textbf{3},   is $Sp(2n,\R)$ or $SO(s,t)$ for $\hat c$ odd,  respectively, even; that is, the $tt^*$ group
$G(\R)$ coincides with the VHS automorphism group (cfr.\! eqn.\eqref{groups}).
Moreover there is a $U(1)$ grading element $Q\in\mathfrak{g}\otimes \C$ such that\footnote{\ The adjoint action of $Q$
on $\mathfrak{g}$ gets transported on the bundles $\mathscr{V}_q\to M$ because these bundles are the pull-back (\emph{via} the period map) of homogeneous bundles on the Griffiths domain. See, e.g.\! chapter 11 of \cite{book}.} \cite{Cecotti:1992vy} 
\be\label{WQ}
[Q,C]=-C,\qquad Q\big|_{\mathscr{V}_q}=q\;\mathrm{Id}_{\mathscr{V}_q}.\tag{42}
\ee
Refs.\!\cite{Cecotti:1990wz,Cecotti:1991me,simpson}
show that the
VHS geometry of the complex moduli of a CY $d$-fold is described by a $tt^*$ geometry which 
satisfies the additional conditions \eqref{deedccc}-\eqref{WQ}. The Lie sub-group $H\subset G$
(cfr.\! eqn.\eqref{WH})
 is the centralizer of the $U(1)$ charge operator $Q$ in $G$.\footnote{\ Here it is crucial that in the SCFT case the real Lie group $G$ is of ``Mumford-Tate type'' i.e.\!
 that it contains a \emph{compact} maximal torus, i.e.\! $\mathrm{rank}\,G=\mathrm{rank}\, K$ \cite{swampIII}.}

\subsubsection*{5. Proof of eqn.\eqref{torsionless}}

The crucial fact is that a solution to $tt^*$ corresponds to a pluri-harmonic map 
\be
\underline{\varpi}\circ\underline{E}\colon M\to \Gamma\backslash G/K,\tag{43}
\ee
hence, in particular, to a solution of the $\sigma$-model \eqref{sigmamode}. Since $\underline{\varpi}\circ\underline{E}$ is a solution for \emph{all} K\"ahler metrics on $M$ (cfr.\! \S.\,\textbf{2}), the $tt^*$ map $\underline{\varpi}\circ\underline{E}$ is in particular harmonic for the WP metric $G_{j\bar k}$.

The particular $\sigma$-model solution which describes the moduli geometry of a CY moreover has \emph{finite} action
in the sense that
\be\label{sigma-action}
\int_M d^{2m}\,\sqrt{G}\, L_{\sigma\text{-model}}<\infty \quad\text{where } L_{\sigma\text{-model}}\equiv
\frac{1}{2}\,G^{\alpha\beta}\,h(\phi)_{ab}\, \partial_\alpha \phi^a\,\partial_\beta \phi^b\tag{44}
\ee 
We defer the proof of \eqref{sigma-action} to \S.\,8 below.
 
This shows eqn.\eqref{torsionless}. The argument works for all dimensions $d$ of the Calabi-Yau.

\subsubsection*{6. Review of \cite{Cecotti:1992vy}}

The Hodge metric $K_{j\bar k}$ was introduced in $tt^*$ geometry in ref.\!\cite{Cecotti:1992vy}, and further studied in \cite{Bershadsky:1993ta},
for its relation with the $\tau$-function of isomonodromic problems and, respectively, the Ray-Singer torsion. 
 As already mentioned, in VHS theory the Hodge metric
 makes sense in the complex moduli space of \emph{any} projective variety, Calabi-Yau or not.
Correspondingly, from a $tt^*$ perspective the Hodge metric should be a good K\"ahler metric for all 2d (2,2) QFTs whether they are 
superconformal or not. When the 2d theory is superconformal, however, the metric $K_{j\bar k}$ (restricted to 
the exactly marginal deformations)
 has nicer properties.
\medskip

For a general (2,2) QFT the Hodge metric reads \cite{Cecotti:1992vy}
\be\label{taum}
K_{j\bar k}= \mathrm{tr}\big[C_j \,\bar C_{\bar k}\big],\tag{45}
\ee
where $C_j$ and $\bar C_{\bar k}$ are the coefficients of the matrix-valued 1-forms $C\equiv C_j\, dt^j$ and 
$\bar C\equiv \bar C_{\bar k}\, d\bar t^{\bar k}$ (cfr.\! \eqref{eopro}); $\{t^j\}$ are complex coordinates in the
parameter space 
$P$ of the (2,2) QFT 
\cite{Cecotti:1991me}. 

In the superconformal case we restrict the 1-forms $C$, $\bar C$ to the conformal submanifold
$M\hookrightarrow P$, i.e.\! to marginal deformations.
Conservation of the conformal $U(1)$ charge yields
\be\label{stuoideq}
\mathrm{tr}[C_i\, C_j]=0,\tag{46}
\ee
and  we can rewrite equation \eqref{taum} in arbitrary \emph{real} (that is, not necessarily holomorphic) local coordinates $x^\alpha$ in the form
\be\label{ooza1}
\begin{split}
ds^2_\textsc{Hodge}\equiv K_{\alpha\beta}\,dx^\alpha\,dx^\beta&= \mathrm{tr}\big[(C+\bar C)_\alpha (C+\bar C)_\beta\big]dx^\alpha\,dx^\beta=\\
&=\mathrm{tr}\big[(S^{-1}\partial_\alpha S)^{\rm o} (S^{-1}\partial_\beta S)^{\rm o}\big]dx^\alpha\,dx^\beta,
\end{split}\tag{47}
\ee
where, in the second line, we used eqn.\eqref{eopro}. From eqn.\eqref{emtens} we see that the energy-momentum tensor of the $\sigma$-model, evaluated on the particular $tt^*$ on-shell field configuration $E=S$, is
\be\label{WT}
T_{\alpha\beta} = K_{\alpha\beta}-\frac{1}{2}\, G_{\alpha\beta}\, G^{\gamma\delta}\,K_{\gamma\delta}.\tag{48}
\ee

In refs.\!\!\cite{Cecotti:1992vy,Bershadsky:1993ta} there is a second formula for the Hodge metric -- this one valid only
along the conformal manifold $M$ of a  superconformal $tt^*$ geometry. It is convenient to
adopt the holomorphic gauge \eqref{hologauge}.
We write $g_{a\bar b}$ for the $tt^*$ Hermitian metric on the fibers of the vacuum bundle $\mathscr{V}$ 
written in a holomorphic trivialization, and
$(g_q)_{u\bar v}$ for its restriction to the sub-bundle $\mathscr{V}_q\subset \mathscr{V}$ 
of definite $U(1)$ charge $q$, cfr.\!\! \eqref{deedccc},\eqref{WQ}. 
For the sub-bundle $\mathscr{V}_{1-\hat c/2}\cong T\mspace{-1.5mu}M$ we use the 
holomorphic local frame $\{\partial_{z^j}\}$ with $z^j$ complex coordinates on $M$; from now on indices from the middle of the latin alphabet $j,k,l,\dots$
 always refer to tensors defined in this holonomic holomorphic trivialization of $\mathscr{V}_{1-\hat c/2}\subset\mathscr{V}$.
Then, along the submanifold $M\hookrightarrow P$, one has \cite{Cecotti:1992vy,Bershadsky:1993ta}
\be\label{kappot}
K_{j\bar k}=\partial_i\bar \partial_{\bar k}\mspace{-4mu}\left(\sum_{{q<0}}2q \log \det g_q \right)\!\!.\tag{49}
\ee
Eqn.\eqref{kappot} says that the Hodge metric is
 the curvature of the Griffiths canonical bundle \cite{griIII}
\be
\bigotimes_{q<0} \big(\det \mathscr{V}_q\big)^{-2q}\to M\qquad\text{(note that $2q\in \Z$)}\tag{50}
\ee
equipped with its canonical Chern connection \cite{periods}.
For comparison, the WP metric is the curvature of the line bundle $\mathscr{V}_{-\hat c/2}\to M$
\cite{Cecotti:1991me,stro}
\be\label{WPmm}
G_{j\bar k}= -\partial_j\bar\partial_{\bar k} \log g_{-\hat c/2}.\tag{51}
\ee

The $tt^*$ equations yield a simple formula for the Riemann tensor of the WP metric
on the conformal manifold of a (2,2) SCFT. Taking the trace, we get a universal formula for
the Ricci tensor valid on $M$ for all $\hat c$
\begin{equation}\label{rrricc}
R_{j\bar k}= {(C_j\bar C_{\bar k})_l}^l-(m+1)G_{j\bar k},\tag{52}
\end{equation}
where 
\be
m\equiv \dim_\C M\equiv \mathrm{rank}\, \mathscr{V}_{1-\hat c/2},\tag{53}
\ee 
In the special case $\hat c=3$ eqn.\eqref{rrricc} is sometimes called
the `Strominger formula' \cite{stro}.
\medskip

Writing $P_q$ for the orthogonal projection $\mathscr{V}\to\mathscr{V}_q$, we have
\be\label{WF}
\bar\partial_{\bar k}\partial_j\log g_q=-\mathrm{tr}\big(P_q\, \bar\partial_{\bar k}(g\partial_j g^{-1})\big)= -
\mathrm{tr}\big(P_q[C_j,\bar C_{\bar k}]\big), \tag{54}
\ee
where in the last equality we used eqn.\eqref{lax2} in the form\footnote{\ Recall that $D\equiv d+(g\partial g^{-1})$ is the $tt^*$
Chern connection \cite{Cecotti:1991me}.}
\be
0\equiv \mathsf{coeff}\,\zeta^0\,\mathsf{in}\!\left[\big(\boldsymbol{\nabla}^{(\zeta)}\big)^2\Big|_{(1,1)\text{-part}}\right]= D\bar\partial+\bar\partial D+C\wedge \bar C+\bar C\wedge C.\tag{55}
\ee
Setting $q=-\hat c/2$ in \eqref{WF} we recover the formula \eqref{WPmm}.
The next case, $q=1-\hat c/2$, yields
\be\label{IA}
\bar\partial_{\bar k}\partial_j \log g_{1-\hat c/2}=G_{j\bar k}-{(C_j\bar C_{\bar k})_l}^l=
-R_{j\bar k}-m\,G_{j\bar k}.\tag{56}
\ee
The same result may be obtained more directly by the first equation in \eqref{xrqe}
\be
G_{j\bar k}=\frac{(g_{1-\hat c/2})_{j\bar k}}{g_{-\hat c/2}}\quad \Rightarrow\quad
\log \det G= \log \det g_{1-\hat c/2}- m\log g_{-\hat c/2},\tag{57}
\ee
 using the general K\"ahler identity $R_{j\bar k}=-\partial_j\bar\partial_{\bar k}\log\det G$, and eqn.\eqref{WPmm}.
 \medskip
 
From eqns.\eqref{kappot},\eqref{WPmm} and \eqref{IA} we read
 the linear relations between the three tensors $G_{j\bar k}$, $K_{j\bar k}$, and $R_{j\bar k}$ on the moduli space
of a Calabi-Yau $d$-fold. From the general discussion in \S.\,\textbf{0} we know that there are \emph{two} linear relations for 
$d=1,2$ and \emph{one} for $d=3,4$:
\begin{align}
d&=1 && K_{j\bar k}=G_{j\bar k}\tag{58} && R_{j\bar k}=-2\,G_{j\bar k}\\
d&=2 && K_{j\bar k}=2\,G_{j\bar k}\tag{59}&& R_{j\bar k}=-m\,G_{j\bar k}\\
d&=3 && K_{j\bar k}=(m+3)\mspace{1mu}G_{j\bar k}+R_{j\bar k}\label{hodgeWP}\tag{60}\\
d&=4 && K_{j\bar k}=(2m+4)\mspace{1mu}G_{j\bar k}+2R_{j\bar k}\tag{61}.
\end{align}
The first 3 lines are known to mathematicians \cite{guy} (eqn.(60) was first derived in \cite{Cecotti:1992vy,Bershadsky:1993ta}).

\subsubsection*{7. Proof of eqn.\eqref{eisteinequation}}

By eqn.\eqref{stuoideq}
 we may rewrite the linear relations between the tensors in arbitrary real local coordinates $x^\alpha$ since all three tensors have pure type $(1,1)$.
 \medskip

For $d=1$ eqn.(58) implies (cfr.\! \eqref{emtens})
 \be\label{Q1}
 T_{\alpha\beta}\equiv K_{\alpha\beta}-\frac{1}{2}\,G_{\alpha\beta}\,G^{\gamma\delta}\,K_{\alpha\beta}=0\tag{62}
 \ee
 which is the classical Virasoro constraint of the Polyakov world-sheet string action.
 \medskip
 
 For $d=2$ the two equations (59) yield
 \be\label{Q2}
 R_{\alpha\beta}-\frac{1}{2}G_{\alpha\beta}R-\kappa^2\, T_{\alpha\beta} =(m-1)(m+2\kappa^2)G_{\alpha\beta}\tag{63}
 \ee
for all choice of $\kappa^2$. 
\medskip

For $d=3,4$ eqns.(60),(61) give
 \be\label{Q3}
T_{\alpha\beta}\equiv K_{\alpha\beta}-\frac{1}{2} G_{\alpha\beta}\, G^{\gamma\delta} K_{\gamma\delta}
=\gcd(d,2)\!\left(R_{\alpha\beta}-\frac{1}{2} G_{\alpha\beta} R\right)-(m-1)\big(\gcd(d,2)m+d\big)G_{\alpha\beta}.
\tag{64}
\ee
Eqns.\eqref{Q1},\eqref{Q2}, and \eqref{Q3} yield the various cases of \eqref{eisteinequation}. 

\subsubsection*{8. Finite actions and finite volume}

We have two distinct \emph{finite action} statements. First, the $\sigma$-model action \eqref{sigma-action}
is finite, that is, the moduli-space scalar field configuration, seen as a smooth map from $M$ to the target space
$\Gamma\backslash G(\R)/K$, has \emph{finite energy} in the sense of differential geometry. This holds for all CY dimension $d$.
The second statement, valid for $d\leq4$, is that the \emph{total} gravity $+$ matter action \eqref{action} is finite and proportional to the volume of the moduli $M$,
so that the finite volume condition should be re-interpreted as a finite action requirement.  

The $\sigma$-model Lagrangian $L_{\sigma\text{-model}}$ evaluated on the particular  solution describing the CY moduli geometry
 is just
the WP trace of the Hodge metric (cfr.\! eqns.\eqref{taum},\eqref{ooza1}),  a quantity which is bounded by a constant outside a compact $U\Subset M$, as we see
from the known asymptotics of VHS at infinity \cite{griIII}. This shows the first statement.

For the relevant cases $d\leq 4$ we give
an alternative and more direct argument. The cases $d=1,2$ are trivial, so we focus on $d=3,4$.
From eqns.(60),(61)  we have
\be\label{poq1z}
L_{\sigma\text{-model}}\big|_{tt^*\,\text{solution}}\equiv G^{\bar k j}K_{j\bar k}=\left(\frac{m}{\kappa^2}+d\right)\!m+\frac{1}{2\,\kappa^2}R,\tag{65}
\ee
 where $m=\dim_\C M$ and
$R$ is the moduli scalar curvature which is negative outside some compact $U$ \cite{OoV}. Thus, outside the \emph{compact} $U$,
$0\leq L_{\sigma\text{-model}}< (m+3)m$ and
\be
\text{$\sigma$-model action}< \int_{U} d^{2m}z\,\det(G_{j\bar k})\, L_{\sigma\text{-model}} + \left(\frac{m}{\kappa^2}+d\right)\!m\cdot\mathsf{vol}(M\setminus U)<\infty,\tag{66}
\ee
since the volume of $M$ is finite.

In the total action \eqref{action}, the Einstein term $-R/(2\kappa^2)$ cancels the last term in eqn.\eqref{poq1z}
so that 
\be\label{kasqwzz}
\text{total action}=\int_M d^{2m}x\,\sqrt{G}\,\left(\frac{m}{\kappa^2}+d\right) = \left(\frac{m}{\kappa^2}+d\right)\cdot \mathsf{vol}(M),
\tag{67}
\ee
so that WP volume and total action agree up to overall normalization. In particular, since WP volumes of CY moduli spaces are finite
\cite{cymoduli2},
the total action is finite.

\begin{rem} The total action is given by eqn.\eqref{kasqwzz} for all classical solutions of the model \eqref{action}, not just for the particular one describing the moduli WP geometry.
\end{rem}

\subsubsection*{9. The canonical K\"ahler metrics for arbitrary $d$}

Consider the functions on $M$ of the form ($d\equiv \hat c$)
\be\label{pppax}
K^{(c)}\overset{\rm def}{=}\sum_{q=-d/2}^{d/2} c(q)\,\log\det g_q,\qquad c(q)\in\R.\tag{68}
\ee
In a holomorphic gauge where $\det \eta=1$ (they exist \cite{dubro1})
the reality constraint \cite{Cecotti:1991me} implies 
\be
\log \det g_q=-\log\det g_{-q},\tag{69}
\ee
so the function \eqref{pppax} depends only on the $[(d+1)/2]$ combinations $\{c(q)-c(-q)\}_{q<0}$,
the same number as the dimension of the convex cone $\mathscr{C}_d$ of canonical metrics,
and we are free to assume $c(q)$ to be an \emph{odd} function.
The K\"ahler potential of the WP metric has the form \eqref{pppax} with $c(q)=\pm 1/2$ for $q=\pm d/2$ and zero otherwise.
The K\"ahler potential of the Hodge metric has this form with $c(q)=q$, see eqn.\eqref{kappot}.

The general ``God-given'' K\"ahler metric in the cone $\mathscr{C}_d$ is then
\be\label{jasqwer}
\begin{split}
K^{(c)}_{j\bar k}&\equiv\partial_j\partial_{\bar k}\!\left(\sum_{q=-d/2}^{d/2} c(q)\,\log\det g_q\right)=\mathrm{tr}\Big[\big(c(Q+1)-c(Q)\big)C_k\bar C_{\bar k}\Big]\equiv\\
&\equiv
\sum_q \Big(c(q+1)-c(q)\Big)\mathrm{tr}(P_q\,C_j\bar C_{\bar k})\equiv-\mathrm{tr}\Big(\big[c(Q),C_j\big]\bar C_{\bar k}\Big),\end{split}\tag{70}
\ee
where, as before, $P_q$ is the projector $\mathscr{V}\to\mathscr{V}_q$.
$K^{(c)}_{j\bar k}$ is a positive K\"ahler metric when the coefficient function $c(q)$ belongs to the
appropriate convex cone $\mathscr{C}_d\subset \R^{[(d+1)/2]}$
 which manifestly includes the cone of increasing functions $c(q+1)> c(q)$.
 \medskip
 
 We show that all metrics $K^{(c)}_{j\bar k}$ satisfy the \emph{``Bianchi identity''} \eqref{Bianchi}. This property is automatic since these tensors correspond to $\partial\overline{\partial}$-exact (1,1)-forms $\kappa^{(c)}$. $\partial\kappa^{(c)}=0$ reads
 \be
 D_i K^{(c)}_{j\bar k}= D_j K^{(c)}_{i\bar k}\qquad\quad \begin{smallmatrix}\text{$D_i$: the (1,0) part of the\phantom{ma}}\\
 \text{WP Levi-Civita connection}\end{smallmatrix}\tag{71}
 \ee
Contraction with $G^{i\bar k}$ yields the  ``Bianchi identity''.

\subsubsection*{10. The explicit ``Einstein equation'' for arbitrary $d$}

From eqn.\eqref{IA} we get 
\be\label{qwerqtr}
R_{j\bar k}-\frac{1}{2}\, G_{j\bar k}\, R-(1-m^2)\,G_{j\bar k}= \mathrm{tr}\big(P_{1-d/2}\mspace{2mu} C_j\bar C_k\big)-
G_{j\bar k}\, G^{k\bar l}\,\mathrm{tr}(P_{1-d/2}\mspace{2mu} C_k\bar C_l\big)\tag{72}
\ee
which is the explicit form of the linear relation \eqref{EEE}. The tensor in the \textsc{rhs} is conserved by the ``Bianchi identity''. While the \textsc{rhs} looks as a valid energy-momentum tensor when evaluated
on the on-shell configuration $S$, it is hard to find an off-shell action with positive kinetic terms
which reproduces it. Our feeling is that it does not exist.

\section*{Acknowledgment}

It is a pleasure to thank Cumrun Vafa for discussions, suggestions on the manuscript, and for  a long time,
 fruitful collaboration which originated most of the ideas used in this note.

\end{document}